\documentstyle[12pt]{article}
\newcommand{\ba}{\begin{array}}
\newcommand{\ea}{\end{array}}
\newcommand{\bc}{\begin{center}}
\newcommand{\ec}{\end{center}}
\newcommand{\be}{\begin{equation}}
\newcommand{\ee}{\end{equation}}
\newcommand{\hsp}{\hspace{1cm}}
\newcommand{\dd}{\varphi}
\newcommand{\w}{\psi}
\newcommand{\la}{\lambda}
\newcommand{\al}{\alpha}
\newcommand{\bb}{\beta}

\newcommand{\mol}[1]{Mol. Phys. {\bf #1}}

\newcommand{\fbs}[1]{Few-Body Systems {\bf #1}}

\newcommand{\veta}{\vec{\eta}}

\newcommand{\vxi}{\vec{\xi}}
\newcommand{\vr}{\vec{r}}

\newcommand{\rij}{r_{ij}}

\newcommand{\esp}{\vspace{.1cm}}
\newcommand{\beq}{\begin{equation}}
\newcommand{\eeq}{\end{equation}}
\newcommand{\beqa}{\begin{eqnarray}}
\newcommand{\eeqa}{\end{eqnarray}}
\newcommand{\bdis}{\begin{displaymath}}
\newcommand{\edis}{\end{displaymath}}
\begin{document}
\vspace{2.in}
\begin{center}
\begin{Large}
\begin{bf}
Analytical Expressions for a Hyperspherical Adiabatic Basis \\
Three Particles in 2 Dimensions \\
\end{bf}
\end{Large}
\vspace{.7cm}
Anthony  D. Klemm \\
School of Computing \& Mathematics, Deakin University, Geelong, Victoria,
Australia \\
{\it and} \\
Sigurd Yves Larsen \\
Department of Physics, Temple University, Philadelphia, PA 19122, USA  \\
Physics Department, University of South Africa 0003, Pretoria 0003, 
South Africa \\

\vspace{1.cm}
\begin{bf}
ABSTRACT \\
\end{bf}
\vspace{.3cm}
\end{center}
For a particular case of three-body scattering in $2$ dimensions,
we demonstrate analytically that the behaviour
of the adiabatic potential is different from that of the hyperspherical
coupling matrix elements, thereby leading to a phase shift that tends to
zero, as the energy goes to zero, instead of to a constant.
We consider two particles interacting with binary repulsive step potentials,
one acting as a spectator, and  solve analytically for the adiabatic
eigenvalues and eigenvectors, for all values of $\rho$.
We are thereby able to obtain the leading terms in the long range
behaviour of the effective potentials, and confirm its inverse logarithmic 
nature, suggested from a previous numerical study.
\newpage
\topmargin=0in
\pagestyle{plain}
\pagenumbering{arabic}
\setcounter{page}{2}
\begin{center}
\section*{Introduction}
\end{center}
\vspace{.3cm}

In a previous paper\cite{klemm}, the authors show how, starting from
hyperspherical harmonic expansions, they obtained
adiabatic potentials, suitable for the calculations of three-body
phase shifts at low energies. The calculations were for 3 particles
in a plane, subject to finite repulsive core interactions.

\esp
The calculations were meant to establish a method which would lead to
the evaluation, at low temperature, of a third fugacity coefficient
in Statistical Mechanics.
The latter task was subsequently carried out by Jei Zhen
and one of the authors\cite{zhen}.
In both investigations, it was important to consider different
cases, corresponding to the distinct representations of the permutation group
and different physical situations, with either the 3 particles interacting or
simply two of them interacting, with the third acting as a spectator.

\esp
Absolutely crucial, in these investigations, is the large-$\rho$ behaviour
of the adiabatic potentials. The nature of the long ``tail'' of the adiabatic
potential determines how the correspondent eigenphase shift behaves,
as the energy tends to zero. Thus, our most significant result was that for
the 3 most important types of the phase shifts,
associated with the cases of ${}^{0}\Gamma_{1g}$, ${}^{0}\Gamma_{2g}$
and $\overline{\delta}$, the adiabatic potentials (the adiabatic eigenvalue
minus a centrifugal term) behave as $1/(\rho^2 ln \rho)$, for large values of
$\rho$, instead of the $1/\rho^2$ of the hyperspherical potential matrix
elements.
This then implies that the phase shifts, instead of tending to constants
as the energy goes to zero, behave as $1/(\ln\,q)^2$, and therefore
go to zero! (The variables $\rho$ and $q$ are, respectively, the hyper
radius and the reduced wave number.)

\esp
Though, in our old paper, our basic material was numerical,
we were able nevertheless to propose ``heuristic'' formulae, to characterize
the asymptotic behaviour of the $3$ types of eigenpotentials, of the remodeling
that takes place to yield a different scattering from the one expected from
the solution of a finite number of hyperspherical equations.

\esp
In this paper, we present analytical results, where we show that in one of
the three cases mentioned above, the case $\overline{\delta}$, we succeeded
in calculating analytically the adiabatic eigenvectors and eigenvalues for all
the values of $\rho$, and therefore also in the asymptotic region.

\esp
While this calculation involves a case where only two of the particles
interact, while the 3rd particle acts as a spectator, it is well to note that
in the hyperspherical coordinate system the two-body interaction is
long ranged (in $\rho$) and also that in the full hyperspherical
calculations
of the other cases, we only need, using symmetry and enforcing a restriction
on the quantum numbers, to
take into account the matrix element of one of the pair potentials.

\esp
Here, then, our calculations allow us to re-examine our previous results, and
confirm and extend the asymptotic form (and coefficients) that can be used
to characterize the long range behaviour of the adiabatic potentials.

\newpage
\begin{center}
\section*{The KL Hyperspherical Coordinate System}
\end{center}
\subsection*{The Harmonic Basis}
\vspace{.2cm}

For a system of three equal mass particles in two dimensions, 
 we define the Jacobi coordinates
\[
 \veta = (\vr_1 - \vr_2)/\sqrt{2} \hsp \mbox{and} \hsp \vxi = 
 \sqrt{2/3}\, (\frac{\vr_1+\vr_2}{2} - \vr_3)\;,
\]
which allows us to separate, in the Hamiltonian, the center of mass
coordinates from those associated with the internal motion.

\esp
Kilpatrick and Larsen\cite{john} then introduce hyperspherical coordinates,
associated with the moment of inertia ellipsoid, of the $3$ particles,
which allows them to disentangle permutations from rotations and obtain
harmonics which are pure representations of both the permutation
and the rotation group. Taking the $z$ axis normal to the plane of
the masses, we write for the cartesian components of the Jacobi coordinates
\beqa
\eta_{x} & = & \rho (\cos \vartheta \cos \varphi \cos  \psi
                    + \sin \vartheta \sin \varphi \sin \psi ),  \nonumber \\
\eta_{y} & = & \rho (\cos \vartheta \cos \varphi \sin  \psi
                    - \sin \vartheta \sin \varphi \cos \psi ),  \nonumber \\
 \xi_{x} & = & \rho (\cos \vartheta \sin \varphi \cos  \psi
                    - \sin \vartheta \cos \varphi \sin \psi ),  \nonumber \\
 \xi_{y} & = & \rho (\cos \vartheta \sin \varphi \sin  \psi
                    + \sin \vartheta \cos \varphi \cos \psi ),
\eeqa
in terms of
the hyper radius $\rho$ and of the three angles $\vartheta$, $\varphi$ and
$\psi$.

\esp
The harmonics, in their unsymmetrized form, are then 
\be
Y_N^{\nu \la}(\Omega) = C_n^{\al \bb} \Theta_n^{\al \bb}(x) e^{i\nu\dd}
                        e^{i\la\w} 
\ee
   where $x = \sin{2\vartheta}$ and
\be
  \Theta_n^{\al \bb}(x) = (1 - x)^{\al/2}(1 + x)^{\bb/2} P_n^{\al,\bb}(x)
\ee
$P_n^{\al,\bb}(x)\;$ is a Jacobi polynomial, and the
 normalization constant is
\[
 C_n^{\al \bb} = \left\{\left(\frac{N + 1}{2^{\al +\bb + 1}}\right)
              \left( \ba{c} n + \al + \bb \\  \al \ea \right)
              \left( \ba{c} n + \al \\ \al \ea \right)^{-1} \right\}^{1/2}\;.
\]
   The hyper radius $\rho\;$ satisfies $\rho^2 = \eta^2 + \xi^2$
and the angular components have the ranges
\[
 -1 \leq x \leq 1, \hsp -\pi/2 \leq \dd \leq \pi/2,
 \hsp 0 \leq \w \leq 2\pi\;.
\]
   Finally we have for the indices the relations
\[
 n = \frac{1}{2}[N - \mbox{max}\{| \nu|,|\la|\}], 
 \hsp \al = \frac{1}{2}|\nu +\la|, \hsp \bb = \frac{1}{2}|\nu - \la|\;,
\]
   where $N$ is the degree of the harmonic, and $\la$ is the inplane
   angular momentum quantum number. The indices 
 $\nu$ and $\la$ take on the values $-N$ to $N$ in steps
of $2$; all three have the same parity and $N = 0,1,2,\ldots.$

\esp
Linear combinations of the basic harmonics can then be formed\cite{john} 
to obtain irreducible bases, adapted to the symmetries of the
physical problems\cite{klemm,zhen}.
\newpage
\subsection*{The Adiabatic Basis}
\vspace{.2cm}

For our model, the particles interact via a binary step potential
\be
     V(\rij)  =  \left\{ \ba{ll}
                          V_0, & \rij \leq \sigma \\
                          0, & \rij > \sigma
                       \ea
                 \right. \hsp  
\ee
where the height $V_0$, and the range, $\sigma$, are both finite.

\esp
The adiabatic eigenfunctions $B_{l}$ are then defined as satisfying
\be
 \left\{ - \frac{1}{\rho^2}\nabla_{\Omega}^2 +
     \frac{2m}{\hbar^2} V(\rho,\Omega) \right\} B_{l}
     = \lambda_{l}(\rho) B_{l} \;, 
\ee
where $ V(\rho,\Omega)$ is either the sum of the binary potentials or,
simply one of the binary potentials, say $V(r_{12})$, expressed as a function 
of $\rho$ and the angles. The index $l$ stands for the set of quantum
numbers which characterize and index the particular class of solutions.
$\lambda_{l}(\rho)$ is the eigenvalue, which upon subtraction of a 
``centrifugal'' type term yields the effective potential, of concern to
us later on.

\esp
The eigenfunctions may now be used to expand the wavefunctions of
the physical systems:
\be
\Psi = \sum_{l^{\prime}} B_{l^{\prime}}(\rho,\Omega)
\phi_{l^{\prime}}(\rho) \;, 
\ee
where the amplitudes $\phi_{l}(\rho)$ are the solutions of the
coupled equations:
\beq
-\sum_{l^{\prime}}\int\! d\Omega \,B^{\ast}_{l}
(\Omega,\rho)\frac{\partial
^{2}}{\partial\rho^{2}}\left(B_{l^{\prime}}(\Omega,\rho)\phi_{l^{
\prime}}(\rho)\right) + \lambda_{l}(\rho)\phi_{l}(\rho)
= (2mE/\hbar^{2})\phi_{l}(\rho) .
\eeq

The adiabatic eigenfunctions can themselves be expanded in hyperspherical
harmonics and this is how a large set of them were calculated in the 
papers quoted earlier.
The symmetries of the hyperspherical harmonic basis are, of course,
reflected in the solutions of the adiabatic eigenvectors. For the fully
symmetric Hamiltonian, the set of solutions divides into nine separate
subsets\cite{zhen}, each requiring calculations involving combinations of 
matrix elements of only one of the binary potentials, but with restrictions
on the quantum numbers of the unsymmetrized harmonics involved.
In the case of two interacting particles, with a third as a spectator,
we find an additional four subsets. 

\esp
The numerical approach was then, for each $\rho$, to evaluate a large 
potential matrix, with the appropriate harmonic basis, add to this the
(diagonal) ``centrifugal'' contribution arising from the angular part of
the kinetic energy (the angular part of the Laplacian in the Hamiltonian)
and diagonalize to obtain the required adiabatic eigenvalues. The
number of harmonics, needed for numerical convergence, increases as a function
of $\rho$, but it was our fortunate experience to find that it was possible
to evaluate correctly the eigenvalues, that we sought, for values of
$\rho$ large enough that the behaviour of $\lambda_{l}(\rho)$ could
be described by asymptotic forms. We were able to characterize them, and 
this gave us the values of $\lambda_{l}(\rho)$ for all the larger
values of $\rho$.

\newpage
\begin{center}
\section*{Dual Polar Set of Coordinates}
\end{center}
\subsection*{The Harmonic Basis}
In this part of the paper we wish, exclusively, to consider the case of
two particles interacting together, the third acting as a spectator.
As we shall show, we are then able to obtain exact adiabatic solutions.

Our reasoning is as follows.
When the third particle does not interact with the other two, this must 
imply that the motion of the pair (1,2), and therefore its angular momentum,
is unaffected by the motion of the third particle. In a parallel fashion,
the motion of the third particle, and its angular momentum about the center
of mass of the particles (1,2), must be a constant as well.
If we choose our coordinates carefully, the angular behaviour of two of the
angles should ``factor'' out and, for a given $\rho$, only one variable should
be involved in a key differential equation.

We note that in the KL coordinates, the distances between particles involve
two of the angles, for example $r_{12}^2$ equals $\rho^2 (1 + \cos 2\vartheta
\cos 2\varphi)$.  To get around this, we choose an angle to give us the 
ratio of the length of the $2$ Jacobi vectors, and then polar coordinates
for each of them.
Thus, we represent $\Omega$ by $(\theta_1, \theta_2, \phi)$, where 
 $\eta = \rho \cos \phi$, $\xi = \rho \sin \phi$ and $\eta_x = \eta \cos
\theta_1$,  $\eta_y = \eta \sin \theta_1$,  $\xi_x = \xi \cos \theta_2$,
 $\xi_y = \xi \sin \theta_2$. The ranges of these angles are
\[
   0 \leq \phi \leq \pi/2, \hsp 0 \leq \theta_1 \leq 2\pi, \hsp 0 \leq
\theta_2 \leq 2 \pi \;.
\]

To obtain the harmonics, in a manner which is suitable to also demonstrate the 
link with the KL harmonics, we introduce complex combinations of the Jacobi
coordinates, i.e. the monomials
\beqa
z_{1} & = &  (\eta_{x} + \imath \eta_{y}) + \imath (\xi_{x} + \imath \xi_{y})
                     \nonumber \\
z^{*}_{1} & = &  (\eta_{x} - \imath \eta_{y}) - \imath (\xi_{x} - \imath
\xi_{y}) \nonumber \\
z_{2} & = & (\eta_{x} - \imath \eta_{y}) + \imath (\xi_{x} - \imath \xi_{y})
                     \nonumber \\
z^{*}_2 & = & (\eta_{x} + \imath \eta_{y}) - \imath (\xi_{x} + \imath \xi_{y})
\eeqa
It then follows that 
\beqa
\rho^2 & = & \frac{1}{2} (z_{1} z^{*}_{1} + z_{2} z^{*}_{2}) \nonumber \\
\nabla^{2} & = & 8 (\frac{\partial^{2}}{\partial z_{1} \partial z^{*}_{1}}
   + \frac{\partial^{2}}{\partial z_{2} \partial z^{*}_{2}}),
\eeqa
and, clearly, $z_{1}$, $z^{*}_{1}$, $z_{2}$ and $z^{*}_2$ each satisfies
Laplace's equation, as do the combinations $z_{1}z_{2}$, $z_{1}z^{*}_{2}$,
$z^{*}_{1}z_{2}$, $z^{*}_{1}z^{*}_{2}$ and these combinations raised to 
integer powers. 

Writing $\rho_{1}^{2} = z_{1}z^{*}_{1}$
and $\rho_{2}^{2} = z_{2}z^{*}_{2}$, we can write as the most general solution
arising from the monomials $z_{1}$ and $z_{2}$:
\bdis
z_{1}^{l_{1}} z_{2}^{l_{2}} P_{l}^{l_{2},l_{1}}(\frac{\rho_{2}^{2} -
\rho_{1}^{2}} {\rho_{2}^{2} + \rho_{1}^{2}}) ( \rho_{1}^{2} +
\rho_{2}^{2} )^{l},
\edis
where $l_1$, $l_2$ and $l$ are positive integers or zero, and 
$P_{l}^{l_{2},l_{1}}$ is a Jacobi polynomial. \\ 
\newpage
\noindent
In terms of the angles, our expression becomes proportional to:
\bdis
\rho^{l_{1} + l_{2} + 2l } (\cos^{2} \phi)^{l_{1}/2} (\sin^{2} \phi)^{l_{2}/2}
P_{l}^{l_{2},l_{1}}(\cos 2 \phi) e^{\imath\theta_{1}l_{1}} 
e^{\imath\theta_{2}l_{2}}
\edis
and, finally, in terms of $z$ equal to $\cos 2\phi$, we define our 
unnormalized harmonic:
\beq
Y_{l}^{l_{1},l_{2}}(\theta_{1}, \theta_{2}, z) = 
(1 + z)^{|l_{1}|/2}(1 - z)^{|l_{2}|/2}\,
P_{l}^{|l_{2}|,|l_{1}|}
(z)\, e^{\imath\theta_{1}l_{1}} e^{\imath\theta_{2}l_{2}},
\eeq
where now $l_{1}$ and $l_{2}$ can be positive, negative, integers - or zero.
(This takes into account the other combinations $z_{1}z_{2}^{*}$, etc.)
The order of the harmonic is $N$ equal to $|l_{1}| + |l_{2}| + 2l$. 
\vspace{.5cm}
\subsection*{The Adiabatic Differential Equation}

Writing
\beq
\nabla_{\eta}^{2} + \nabla_{\xi}^{2} = (\,\frac{\partial^{2}}
{\partial\rho^{2}} + \frac{3}{\rho}\frac{\partial}
{\partial\rho}\,) + \frac{1}{\rho^2}\nabla_{\Omega}^2\, ,
\eeq
inserting our polar coordinates into the left hand side and changing to our
variable $z$, we find:
\beq
 \nabla_{\Omega}^2 = 
4 (1 - z^2) \frac{\partial^2}{\partial z^2} - 8 z \frac{\partial}{\partial z}
+ \frac{2}{(1+z)}\frac{\partial^2}{\partial\theta_{1}^{2}}
	      + \frac{2}{(1-z)}\frac{\partial^2}{\partial\theta_{2}^{2}}\,\,.
\eeq
If we now write our adiabatic eigenfunctions as
\beq
 B_{N}^{l_1 l_2}(\rho, \Omega) = e^{i l_{1} \theta_1} e^{i l_{2} \theta_2}\,
 (1 + z)^{|l_{1}|/2}(1 - z)^{|l_{2}|/2}\,  F_{l}^{|l_1|,|l_2|}(\rho,z)\;,
\eeq
then the functions $F$ will satisfy the equation:
\beqa
 \left[ - 4 (1 - z^2) \frac{\partial^{2}}{\partial z^{2}}  + 
4((2 + l_1 + l_2)z + l_2 - l_1)\frac{\partial}{\partial z} \right] & \hspace{-3.9cm}F_{l}^{l_1,l_2}(\rho, z) \nonumber \\
+ \left[ (l_1 + l_2)(l_1 + l_2 + 2) +
  \rho^2 \overline{V}(\rho,z) \right] & \!\!\!\!\!F_{l}^{l_1,l_2}(\rho, z)
= 
\rho^2 \lambda(\rho) F_{l}^{l_1,l_2}(\rho, z) 
\eeqa
where $\overline{V}(\rho,z)$ equals $2m/\hbar^{2}$ times the potential and
in our notation we have dropped the absolute value indications.

\esp
When $\overline{V}(\rho,z) = 0$, we can  obtain a 
solution which is analytic between $-1 \leq z \leq +1$. 
For $\lambda$ equal to $ (l_1 + l_2 + 2 l)(l_1 + l_2 + 2l + 2)/\rho^2$ and
$l$ a non-negative integer, we 
find that our $F$ is simply $P_{l}^{l_{2},l_{1}}(z)$, the Jacobi polynomial
which appears in our Eq. (10).
The $N$ that appears in the $B$ of Eq. (13) is the order of the
corresponding harmonic.

\esp
For our potential  
\be
    \overline{V}(\rho,z) = \left\{ \ba{ll}
   (2m/\hbar^2) V_{0} & -1 \leq z \leq -1 + 1/\rho^2 \\
               0  & -1 + 1/\rho^2 < z \leq 1 \;,
                     \ea \right. 
\ee
the solutions of this equation which behave reasonably at 
$z$ equal to $-1$ and $+ 1$ will be seen to be
proportional to extensions of the Jacobi polynomials to functions with
non-integer indices, in a relationship similar to that of Legendre 
polynomials and Legendre functions.
\newpage
To motivate and clarify our
procedure we first consider the case of $l_1 = l_2 = 0$, with and without
potential.

When the potential is put to zero and we factor a $4$ as well as change the 
sign, the differential equation reads
\beq
  \left[(1 - z^2) \frac{\partial^{2}}{\partial z^{2}}  - 2z 
 \frac{\partial}{\partial z} + l\,(l + 1)\right] \; F_{l}^{0,0}(\rho, z) = 0.
\eeq
This is, of course, the Legendre differential equation and, with $l$ a positive
or zero integer, the well behaved solutions are the Legendre polynomials.

In the case of our potential, which is zero or a constant (only a function of
$\rho$) in the different ranges of $z$, we can write our differential equation
in a very similar form, i.e. as
\beq
\left[(1 - z^2) \frac{\partial^{2}}{\partial z^{2}}  - 2z
\frac{\partial}{\partial z} + \nu\,(\nu + 1)\right] \; F_{\nu}^{0,0}
(\rho, z) = 0 \,,
\eeq
where for $-1+1/\rho^2 < z \leq 1$
\beq
\hspace{-1.4cm}\nu\,(\nu + 1) = \rho^2\,\lambda\,(\rho)/4
\eeq
and\, for $-1 \leq z \leq -1+1/\rho^2$
\beq
\nu\,(\nu + 1) = \rho^2\,[ \lambda\,(\rho) - \overline{V}_0]/4 \;.
\eeq

Denoting the respective values of $\nu$ as $\nu_1$ and $\nu_2$, the
corresponding solutions are the Legendre function $P_{\nu_1}(z)$ and the
combination 
\bdis
 P_{\nu_{2}}(-z) = \cos(\pi \nu_{2})\, P_{\nu_{2}}(z) - 
(2/\pi)\, \sin(\pi \nu_{2})\, Q_{\nu_{2}}(z)\,,
\edis
of the first and second Legendre functions.

\esp
The point is as follows. Whereas $P_{\nu_{1}}(z)$ is well behaved at
$z$ equal to $1$, and is suitable as a solution for its
range in $z$ from $-1 + 1/\rho^2$ to $1$, both the 
$P_{\nu_{2}}(z)$ and $Q_{\nu_{2}}(z)$ have a logarithmic singularity at $z$ 
equals $-1$. The combination that we propose, however, is such that the
logarithmic terms cancel out and the combination\cite{abram} is a well
behaved solution in the range  $-1$ to $-1 + 1/\rho^2$.

Expressing these solutions as power series, the first about $z$ = $1$,
the second
about $z$ = $-1$, we obtain 
\be  \ba{cl}
 {}_2F_1(-\nu_1, \nu_1 + 1; 1;
    \frac{1}{2}(1-z)), & \mbox{for} -1+1/\rho^2 < z \leq 1 \nonumber \\
   \mbox{and} & \hsp \\
 {}_2F_1(-\nu_2, \nu_2 + 1; 1;
       \frac{1}{2}(1+z)), & \mbox{for} -1 \leq z \leq -1+1/\rho^2
\ea
\ee
Our overall solutions are then obtained by matching the logarithmic
derivative of the two solutions (above) at the boundary: 
at $z$ equal $-1 + 1/\rho^2$. 
This then also yields the adiabatic eigenvalues.

It now remains to note that for the cases of $l_1$ and $l_2$ not 
equal to zero, we can use the same procedure. We have, for the two regimes,
solutions proportional to 

\be  \ba{cl}
 {}_2F_1(-\nu_1, \nu_1 + |l_1| + |l_2| +1; |l_2| + 1;
    \frac{1}{2}(1-z)), & \mbox{for} -1+1/\rho^2 < z \leq 1 \nonumber \\
   \mbox{and} & \hsp \\
 {}_2F_1(-\nu_2, \nu_2 + |l_1| + |l_2| + 1; |l_1| + 1;
       \frac{1}{2}(1+z)), & \mbox{for} -1 \leq z \leq -1+1/\rho^2
\ea
\ee

\newpage
 For each choice of $l_1$ and $l_2$
there is an infinite set of values of $\nu_1\;$ for which the logarithmic
derivative of the hypergeometric functions can be matched at $z$ equal to 
$-1 + 1/\rho^2$. For each such value of $\nu_1$, the 
adiabatic eigenvalue is then given by
\be
  \lambda (\rho) = \frac{(2 \nu_1 + |l_1| + |l_2| + 1)^2 - 1}
                              {\rho^2} \;. 
\ee


When $V_0 = 0$, the adiabatic basis
reduces to the hyperspherical harmonic basis of Eqn. (10), since
 the hypergeometric functions reduce to Jacobi polynomials, and
 $\nu_1 \equiv \nu_2 = l$. So our $B_{N}^{l_1,l_2}$ is precisely the 
$Y_{l}^{l_1,l_2}(\theta_1,\theta_2,z)$.
\section*{Comparison of the Adiabatic Eigenvalues}

When the numerical work was done (using the KL basis),
lists were made of the appropriate harmonics needed to form the
matrices (potential and centrifugal) which, when added and diagonalized,
yield the adiabatic eigenvalues. We now need to identify these
eigenvalues and compare them with those obtained by the new method. This
is not trivial, but an immediate remark can be made.

First of all, the angular momentum $\la$ is a good quantum number, with 
\be
 \la = l_1+ l_2 \;. 
\ee
 This follows from the fact that $l_1$ specifies the angular momentum of
the 1-2 pair and $l_2$ specifies the angular momentum of the third particle
relative to the center of mass of the first two. Thus their sum defines the
total inplane angular momentum. Hence, for example, when $\la = 0$ we can
have all pairs $l_1$ and $l_2$ with $l_1 = - l_2$. If $l_1 = l_2 = 0$,
this then provides a single eigenvalue for each choice of $N = 2\,l,\,
l = 0,1,2,\ldots.$

Another indicator is wether $n$ is even or odd, which is very significant in
the  drawing up of the lists, associated with the symmetries of the harmonics.
Proceeding, then, we compare values of the effective potential, defined by 
\be
  V(\rho, N) = \la(\rho) - \frac{(N+1)^2-\frac{1}{4}}{\rho^2}\;,
\label{22}
\ee
where we subtract from each eigenvalue the value of the centrifugal term 
that would correspond to it, if the binary potential were allowed to go to
zero.
These have been extensively tabulated by Zhen\cite{jei}.

  {\em Table 1} compares the results in the simplest case, $N = 0$, where we
demonstrate the convergence of the trucated matrix method with the result
obtained directly, for a sample value of $\rho = 5\;$ and $\Lambda^* = 10$.
($\Lambda^* = (h^2/mV_0\sigma^2)^{1/2}$)
\bc
\begin{tabular}{|cc|} \hline
$N_{max}$ & $V(5, 0)$ \\ \hline
 $110$ & $0.011754744$ \\
 $120$ & $0.011754730$ \\
 $130$ & $0.011754670$ \\
 $140$ & $0.011754666$ \\ 
 Direct & $0.011754562$ \\ \hline
\end{tabular}
 {\it Table 1. Convergence of the matrix method}
\ec
\newpage

  A more extensive set of comparisons is made in {\em Table 2}, where selected
values of the effective potential, obtained from eigenvalues of the truncated
matrix, are chosen for various values of $N$, $\la$ and $n$ and compared
with the direct results. In all cases, except the first,
the matrix was truncated at $N_{max} = 100$.
\bc
\begin{tabular}{|llll|llll|} \hline
\multicolumn{8}{|c|}{$V(\rho,N)$} \\ \hline
\multicolumn{4}{|c}{\em Truncated Matrix} & 
   \multicolumn{4}{c|}{\em Direct} \\ \hline
$n$ & $\la$ & $N$ & $V(5,N)$ & $l$ & $|l_1|$ & $|l_2|$ & $V(5,N)$ \\ \hline
E & 0 & 0 & 0.011754666 & 0 & 0 & 0 & 0.011754562 \\
E & 0 & 2 & 0.037577818 & 1 & 0 & 0 & 0.037577462 \\
O & 0 & 2 & 0.000874927 & 0 & 1 & 1 & 0.000874911 \\
E & 0 & 4 & 0.062609805 & 2 & 0 & 0 & 0.062609219 \\
E & 0 & 4 & 0.00005971  & 0 & 2 & 2 & 0.00005971  \\
O & 2 & 4 & 0.00413519  & 1 & 1 & 1 & 0.00413512  \\
E & 1 & 1 & 0.024168    & 0 & 0 & 1 & 0.02416738  \\
E & 1 & 1 & 0.00029592  & 0 & 1 & 0 & 0.00029591  \\
O & 1 & 3 & 0.000024    & 0 & 2 & 1 & 0.00002426  \\
O & 1 & 3 & 0.00172537  & 0 & 1 & 2 & 0.00172529  \\
E & 1 & 3 & 0.050462    & 1 & 0 & 1 & 0.0504588   \\
E & 1 & 3 & 0.00226748  & 1 & 1 & 0 & 0.00226737  \\
E & 2 & 2 & 0.00000616  & 0 & 2 & 0 & 0.00000616  \\
O & 2 & 2 & 0.036849    & 0 & 0 & 2 & 0.03684737  \\
E & 2 & 4 & 0.000088636 & 1 & 2 & 0 & 0.000088629 \\
E & 4 & 4 & 0.062866    & 1 & 0 & 2 & 0.06286247  \\ \hline
\end{tabular}

{\it Table 2. Some effective potential values in the ``delta-bar'' class}
\ec
\section*{Asymptotic Behaviour}
    The matching of logarithmic derivatives provides a means of
 obtaining information about the asymptotic behaviour of the eigenvalues,
 and hence the effective potentials, as the hyper-radius, $\rho$, gets
 large.  
   There is however a particular difficulty in finding this behaviour.
It is that it is \underline{not} simply a case of  looking at the limiting
 behaviour of ${}_2F_1(a, b; c; \epsilon)$ and 
${}_2F_1(a, b; c; 1 - \epsilon)$ as $\epsilon \rightarrow 0$, 
 because the expressions corresponding to
 $a$ and $b$ both depend on $\rho$.

    In the simplest case, corresponding to $l_1 = l_2 = l = 0$, we find
\begin{eqnarray}
   \rho^2 V_{\mbox{eff}}(\rho, N) & = & 4(N + 1) \epsilon_0 \nonumber \\
        & \sim & \frac{1}{{\cal A} + {\cal B} \ln \rho}\;, 
\end{eqnarray}
 where
\be
  {\cal A} = \frac{I_0(\sqrt{2A})}{2(N+1)\sqrt{2A}I_1(\sqrt{2A})} +
            \frac{1}{4(N+1)} \ln 2 
\ee
   and
\be
  {\cal B} = \frac{1}{2(N+1)}  \;. 
\ee
   The $I_i$'s being modified Bessel functions of integer order of the 
first kind.

   The next simplest case is with only $l_1 = 0$. Then
\be
  {\cal A} = \frac{1}{(N+1)} \left[ \frac{I_0(\sqrt{2A})}
 {2\sqrt{2A}I_1(\sqrt{2A})}
         + \frac{1}{4} \ln 2 - \frac{1}{4} \sum_{p=1}^{k} \frac{1}{p} -
         \frac{1}{4} \sum_{q=1}^{k+m} \frac{1}{q} \right] 
\ee
   and
\be
 {\cal B} = \frac{1}{2(N+1)} \;, 
\ee
   where it should be understood that \( \sum_1^0 \equiv 0 \).

    The case with $l_1 \not= 0$ introduces considerable complications,
particularly to the form of $\cal A$, with higher order Bessel functions
occurring, and so is not listed. However the expression for $\cal B$ is
\underline{exactly} the same. This
 confirms the value postulated by Larsen \cite{klemm}.

    It is impressive how well Zhen \cite{jei} did working with
approximate values for the eigenvalues. In her thesis she compares 
her $\cal B$'s with the postulated values. 
If we compare her $\cal A$'s with the above expressions
  (see {\em Table 3}), we see just how consistent her calculations are.

\bc
\begin{tabular}{|lll|c|c|} \hline
$N$ & $l$ & $l_2$ & $\cal A$ (Zhen) & $\cal A$ (here) \\ \hline
 0  &  0  &  0  &  2.6064         &    2.8293   \\
 2  &  1  &  0  &  0.7581         &    0.7764  \\
 4  &  2  &  0  &  0.4146         &    0.4159  \\
 1  &  0  &  1  &  1.2381         &    1.2897   \\
 3  &  1  &  1  &  0.5493         &    0.5511  \\
 5  &  2  &  1  &  0.3356         &    0.3327 \\  \hline
\end{tabular}
 
  {\it Table 3. Comparison of numerical and analytic \\
       asymptotic leading terms.}
\ec
  Thus, the conclusions previously obtained by assuming this form
 of asymptotic behaviour \cite{zhen} are verified; at least in the
 ``delta-bar'' case.

\newpage
\section*{Conclusion}
     It is now clear that the extensive numerical calculations of Zhen
\cite{jei}, using the truncated matrix approach, provided good estimates of
the eigenvalues, the effective potentials, and the $2 + 1$
phase shifts of the third cluster.
 The results are consistent for the entire range of
values of $\rho$, taking into consideration the requirement for larger
$N_{max}$ at larger values of $\rho$.

We were also able to demonstrate the all important logarithmic behaviour 
in the asymptotic form of some of the effective potentials. This insures
that the corresponding phase shifts (dominant at low energies)
go to zero, as the wave number goes to zero.
For the other $2 + 1$ phase shifts, characterized by other 
group classifications of the harmonics, we can demonstrate by explicit
calculations that both the asymptotic form of the effective potentials and
the phase shifts go to zero in a stronger manner.


We would love to obtain similar asymptotic expressions for the effective
potentials of the fully interacting problem. If we were able to do this, it
would simplify enormously the cluster calculations, as well as increase
its accuracy.

\vspace{0.5cm}

\section*{Acknowledgements} This work was supported in part by a Department
of Industry,
Technology and Commerce Bilateral Science and Technology Program (Australia)
 grant and a study grant made available by Deakin University.

\end{document}